\documentstyle[seceq,preprint]{jpsj}

\def\gsim{\,$\raise0.3ex\hbox{$>$}\llap{\lower0.8ex\hbox{$\sim$}}$\,}
\def\lsim{\,$\raise0.3ex\hbox{$<$}\llap{\lower0.8ex\hbox{$\sim$}}$\,}

\def\runtitle{Ground State Phase Diagram of Frustrated $S = 1$ $XXZ$ chains : 
Chiral Ordered Phases}
\def\runauthor{Toshiya {\sc Hikihara}, Makoto {\sc Kaburagi}, 
Hikaru {\sc Kawamura}, Takashi {\sc Tonegawa}}

\title{Ground State Phase Diagram of Frustrated $S = 1$ $XXZ$ chains : 
Chiral Ordered Phases}

\author
{Toshiya {\sc Hikihara}\footnote{E-mail address: 
hikihara@phys03.phys.sci.kobe-u.ac.jp}, 
Makoto {\sc Kaburagi}$^{1}$, Hikaru {\sc Kawamura}$^{2}$ 
and Takashi {\sc Tonegawa}$^{3}$}

\inst
{Graduate School of Science and Technology, Kobe University, Rokkodai, 
Kobe 657-8501 \\
$^1$Faculty of Cross-Cultural Studies, Kobe University, Tsurukabuto, Nada, 
Kobe 657-8501\\
$^2$Faculty of Engineering and Design, Kyoto Institute of Technology,
Sakyo-ku, Kyoto 606-8585\\
$^3$Department of Physics, Faculty of Science, Kobe University, Rokkodai, 
Kobe 657-8501}

\recdate{September 16, 1999}

\abst
{The ground-state phase diagram of frustrated $S = 1$ $XXZ$ spin chains 
with the competing nearest- and next-nearest-neighbor antiferromagnetic 
couplings is studied using the infinite-system density-matrix 
renormalization-group method.
We find six different phases, namely, 
the Haldane, gapped chiral, gapless chiral, double Haldane, N\'{e}el, 
and double N\'{e}el (uudd) phases.
The gapped and gapless chiral phases are characterized by the spontaneous 
breaking of parity, in which the long-range order parameter is a chirality, 
$\kappa _l$=$S_l^xS_{l+1}^y-S_l^yS_{l+1}^x$, whereas 
the spin correlation decays either exponentially or algebraically.
These chiral ordered phases appear in a broad region in the phase diagram for 
$\Delta < 0.95$, where $\Delta$ is an exchange-anisotropy parameter.
The critical properties of phase transitions 
are also studied.}

\kword
{frustration, spin-1 chain, ground-state phase diagram, 
 density-matrix renormalization-group method, chiral ordering}

\begin{document}
\sloppy
\maketitle

\section{Introduction}\label{sec:int}

Frustrated antiferromagnetic quantum spin systems have attracted 
considerable attention over the decades since they exhibit a wealth of 
fascinating phenomena in their ground states and low-lying excitations.
In general, frustration supresses antiferromagnetic correlations and 
the tendency towards the N\'{e}el order.
Classical systems, for example, often show a helical ordered state
in their ground states in the presence of strong frustration.
In quantum systems, the interplay of frustration and quantum fluctuations 
plays an important role which causes exotic phenomena, {\it e.g.}, 
a spin-liquid state and a novel type of spontaneous symmetry breaking.

In this paper, we study a one-dimensional anisotropic spin system with 
the antiferromagnetic nearest-neighbor coupling $J_1$ and the frustrating 
next-nearest-neighbor coupling $J_2$.
The model is described by the $XXZ$ Hamiltonian
\begin{equation}
{\cal H} = \sum_{\rho=1}^{2} \left\{ J_\rho \sum_l \left(
     S_l^x S_{l+\rho}^x + S_l^y S_{l+\rho}^y
    + \Delta S_l^z S_{l+\rho}^z \right) \right\},  \label{eq:Ham}
\end{equation}
where ${\mib S}_l$ is a spin-$S$ spin operator at site $l$ and $\Delta \ge 0$ 
represents an exchange anisotropy.
Hereafter, we put $j \equiv J_2/J_1$ ($j \ge 0$).

In the classical limit, $S \to \infty$, the system exhibits a magnetic 
long-range order (LRO) characterized by a wavenumber $q$.
The order parameter is defined by
\begin{equation}
{\mib m}(q) = \frac{1}{L} \sum_l {\mib S}_l e^{{\rm i}ql}, \label{eq:hel}
\end{equation}
where $L$ is the total number of spins.
While the magnetic LRO is of the N\'{e}el-type ($q = \pi$) 
when $j$ is smaller than a critical value, $j \le 1/4$, 
it becomes of helical-type for $j > 1/4$ with 
$q = {\rm cos}^{-1} \left( -1/4j \right)$.
It should be noticed that both the time-reversal and parity symmetries are 
broken in this helical ordered state.
When the system has an $XY$-like anisotropy ($\Delta < 1$), 
the helical ordered state possesses 
a two-fold discrete degeneracy according as the helix is either right- 
or left-handed, in addition to a continuous degeneracy 
associated with the original $U(1)$ symmetry of the $XY$ spin.
This discrete degeneracy is characterized by mutually opposite
signs of the total chirality defined by\cite{chlOP}, 
\begin{eqnarray}
O_\kappa = \frac{1}{L} \sum_l \kappa_l ,\label{eq:chl} \\
\kappa_l = S_l^x S_{l+1}^y - S_l^y S_{l+1}^x 
         = \left[ {\mib S}_l \times {\mib S}_{l+1} \right]_z .  \nonumber
\end{eqnarray}
Note that this chirality is distinct from the scalar chirality 
of the Heisenberg spin often discussed in the literature\cite{Frahm} 
defined by 
$\chi _l={\mib S}_{l-1}\cdot {\mib S}_l\times {\mib S}_{l+1}$ : 
The chirality $O_\kappa$ changes its sign under the parity operation 
but is invariant under the time-reversal operation while the scalar 
chirality changes its sign under both operations.
Since the frustrated classical chain (\ref{eq:Ham}) always has a planar 
spin order, the scalar chirality $\chi_l$ vanishes trivially.

In the $S = 1/2$ case, the ground-state phase diagram of the system has been 
extensively studied either 
numerically\cite{Hara-Tone,Oka-Nomu1,Oka-Nomu2,Oka-Nomu3} 
or analytically\cite{Maj-Gho,Haldane}.
These studies have revealed that the system undergoes a phase transition 
from the spin-liquid phase to the dimer phase at $j = j_{\rm c}(\Delta)$ 
with increasing $j$.
The critical value $j_{\rm c}(\Delta)$ has been estimated to be 
$j_{\rm c} \sim 0.241$ for the Heisenberg case
($\Delta = 1$)\cite{Oka-Nomu1,Oka-Nomu2,Oka-Nomu3}.
The spin-liquid phase at $j \le j_{\rm c}(\Delta)$ is characterized by gapless 
excitations and an algebraic decay of spin correlations, 
while the dimer phase at $j > j_{\rm c}(\Delta)$ is characterized by a finite 
energy gap above the doubly degenerate ground states and by 
the spontaneous breaking of both the parity and translational symmetries.
The dimer phase is divided into two regions by the Lifshitz line 
$j_{\rm L}(\Delta)$ according to the nature of the short-range spin 
correlations, where $j_{\rm L} = 0.5$ in the Heisenberg case\cite{Tone-Hara}.
Whereas the correlations are of the N\'{e}el type and the structure factor 
$S(q)$ has a maximum at $q=\pi$ for $j \le j_{\rm L}(\Delta)$, 
the correlations become incommensurate and the maximum of $S(q)$ occurs at 
$q = Q < \pi$ for $j > j_{\rm L}(\Delta)$.
This incommensurate character is regarded as the vestige of the helical order 
in the classical case.
It has also been shown both in the $XY$ and Heisenberg cases that 
there is no chiral LRO for arbitrary $j$, i.e., the chirality 
$O_\kappa$ vanishes in the thermodynamic limit\cite{KKH}.

As is well known, on the other hand, the low-energy properties 
of the $S = 1$ spin chain without frustration 
are essentially different from those of the $S = 1/2$ chain. 
Haldane conjectured that integer spin chains are in a phase 
with a finite energy gap in the excitation spectrum 
and with an exponential decay of the two-spin correlation functions, 
while half-odd integer spin chains are in the spin-liquid phase
\cite{Halconj1,Halconj2}. 
This gapped phase for integer spin chains is now called the Haldane phase, 
and is characterized by the string order parameter\cite{Nijs-Rom}
\begin{equation}
O_{\rm str} = \frac{1}{L} \sum_{l} 
\exp \left( \sum_{j=1}^{l-1} {\rm i}\pi S_j^z \right) S_{l}^z. \label{eq:str}
\end{equation}
Haldane's conjecture has been verified either numerically or experimentally, 
and is now well-established.
It is therefore natural to expect that the frustrated $S = 1$ spin chain 
might also exhibit low-energy properties which are substantially 
different from those of both the $S = 1/2$ and the $S = \infty$ 
classical chains.

The aim of this paper is to determine the ground-state phase diagram 
of the model (\ref{eq:Ham}) with $S = 1$ spins in the $j$-$\Delta$ plane.
The ground-state properties of the model have been studied 
in some limiting cases.
In the Heisenberg case ($\Delta = 1$), Kolezhuk {\it et al.} have shown 
that, at $j = 0.7444(6)$, the system undergoes a first-order phase transition 
from the Haldane phase to the ^^ double Haldane' (DH) phase, in which 
the next-nearest-neighbor coupling $J_2$ is dominant and the system can be
regarded as two Haldane subchains coupled with a weak inter-subchain 
coupling $J_1$\cite{Kolezk1,Kolezk2}.
The string order parameter $O_{\rm str}$ vanishes discontinuously 
at the transition.
Meanwhile, in the case of no frustration $j = 0$, 
it has been shown that a phase transition from the Haldane phase 
to the N\'{e}el phase occurs at 
$\Delta = \Delta_{\rm c} \sim 1.18$\cite{Hal-Nel1a,Hal-Nel1b,Hal-Nel2a,Hal-Nel2b}.
The transition in this limit has been verified to belong to 
the universality class of the two-dimensional (2D) Ising model.
It is also known that, in the Ising limit, $\Delta \to \infty$, 
there is a phase transition at $j = 0.5$ between the N\'{e}el phase and 
the ^^ double N\'{e}el' (DN) phase, the latter being characterized by 
the magnetic LRO of ^^ up-up-down-down' (^^ uudd') type.
Systematic studies of the model in the entire $j$-$\Delta$ plane, 
however, has been lacking so far.

In the previous work\cite{KKH}, we have shown that in the pure $XY$ case 
($\Delta = 0$) there occurs a phase transition from the Haldane phase 
to the ^^ chiral ordered' 
phase in which only the chirality exhibits a LRO ($O_\kappa > 0$) with 
no helical LRO (${\mib m}(q) = 0$ for $\forall q$).
Here we note that, in the pure $XY$ and Heisenberg cases, 
the absence of magnetic (spin) LRO has rigorously been proven 
for any $j$ and for general $S < \infty$\cite{Momoi}.
We have also found that there are two types of the chiral phase, 
the ^^ gapped' and ^^ gapless' chiral phases, in which the spin correlation 
decays exponentially and algebraically, respectively.
These chiral phases spontaneously break only the parity symmetry, 
with preserving both the time-reversal and translational symmetries.
The phase defined by this types of symmetry breaking is a novel one 
in quantum systems, while it has been studied quite intensively in various 
frustrated classical systems
\cite{Kawa2,triXY1,triXY2,triXY3,ffXY1,ffXY2,ffXY3,ffXY4,XYSG1,XYSG2,Kawa3}.
It is thus interesting to further investigate the properties of the chiral 
phases in frustrated quantum systems.

In order to determine the phase diagram, we numerically calculate 
the correlation functions associated with the order parameters, ${\mib m}(q)$, 
$O_\kappa$, and $O_{\rm str}$, for open chains, 
using the infinite-system density-matrix 
renormalization-group (DMRG) method.
Examining the behavior of these correlation functions at long distance, 
we find the following six phases , the Haldane, gapped and gapless chiral, 
DH, N\'{e}el, and DN phases, in the region where $j \ge 0$ and $\Delta \ge 0$.
The obtained phase diagram has revealed that the gapped and gapless 
chiral phases exist in a broad region of the diagram where 
$0 \le \Delta < 0.95$.

The plan of this paper is as follows.
In {\S} \ref{sec:DMRG}, we introduce the spin, chiral, and string correlation 
functions associated with each order parameter. 
We also explain our numerical method in some detail.
Results of our numerical calculations are presented in {\S} \ref{sec:res}.
The ground-state phase diagram in the $j$-$\Delta$ plane is constructed 
and the critical properties of the phase transitions 
between different phases are analyzed.
Finally, we summarize our results in {\S} \ref{sec:dis}.

\section{Correlation Functions and Numerical Method}\label{sec:DMRG}

In order to determine the phase diagram, we calculate the spin, chiral, and 
string correlation functions defined by,
\begin{eqnarray}
C_{\rm s}^\alpha (r) = \langle S_{l_0-r/2}^\alpha S_{l_0+r/2}^\alpha \rangle
               ~~~~(\alpha = x,z),  \label{eq:Ccor} \\
C_\kappa (r) = \langle \kappa_{l_0-r/2} \kappa_{l_0+r/2} \rangle ,
                   \label{eq:Cchl} \\
C_{\rm str} (r) = \langle S_{l_0-r/2}^z 
         \exp \left( {\rm i} \pi \sum_{j=l_0-r/2}^{l_0+r/2-1} S_j^z \right)
         S_{l_0+r/2}^z \rangle,      \label{eq:Cstr}
\end{eqnarray}
which are associated with the helical, chiral, and string order parameters, 
respectively.
Here, $l_0$ represents the center position of the open chain, i.e., 
$l_0 = L/2$ for even $r$ and $l_0 = (L+1)/2$ for odd $r$, and 
$\langle \cdots \rangle $ represents the expectation value in the 
lowest energy state in the subspace of $S_{\rm total}^z = 0$ with even parity.
We have checked by exact-diagonalization calculation that the ground state of 
the system indeed belongs to this subspace.
We employ the infinite-system DMRG algorithm 
introduced by White\cite{White1,White2}.
The algorithm is accelerated by using the initial Lanczos vector 
derived from the recursion relation proposed by Nishino and Okunishi
\cite{PWFRG1,PWFRG2}.

Here we note the following two points concerning the properties of 
the infinite-system DMRG.
The first is the dependence of the data on the number of kept states, $m$.
Since no analytic form to extrapolate the data to $m \to \infty$ has been 
established so far, in order to obtain precise results we have no choice 
but to increase $m$ until the data converge.
Generally speaking, the convergence is slow when the energy gap is 
small, while it is fast when the gap is large.
In fact, we have kept up to $m = 400$ states in the vicinity of phase 
boundaries and in the gapless chiral phase, whereas $m = 220$ turns out 
to be large enough in the gapped phases.

Besides, we need to check the $L$-convergence of the correlations.
In the infinite-system DMRG, two spins are added in the center of the chain 
at each step of the algorithm.
We continue the procedure until the correlation functions calculated 
around the center become free from the effects of open boundaries.
In general, more steps are needed if the system has 
longer correlations.
We have performed calculations with up to 3500 steps, which corresponds to 
the system of $L = 7000$, until the $L$-convergence of the data of 
the correlation functions was attained.

\section{Numerical Results}\label{sec:res}
Examining the $r$-dependence of the correlation functions given by 
eqs.(\ref{eq:Ccor})-(\ref{eq:Cstr}), 
we find six different phases in the phase diagram, namely, the Haldane, 
gapped chiral, gapless chiral, DH, N\'{e}el, and DN phases.
Figure~\ref{fig:diag} shows the obtained phase diagram in the 
$j$-$\Delta$ plane.
The long-distance behavior of the correlation functions in each phase 
is summarized in Table \ref{tab:cor}.
In the following subsections,
we discuss our numerical results at each transition line in more detail.

\subsection{Transition to the chiral phases}
Let us first consider the phase transition between the Haldane 
and chiral ordered phases.
In our previous work\cite{KKH}, we studied the phase transition 
in the pure $XY$ case ($\Delta = 0$),
and found that there exist two different types of chiral ordered phase.
In the gapped chiral phase, the chiral and string LRO's coexist and 
the spin correlations decay exponentially with a finite energy gap,
while in the gapless chiral phase, the chiral LRO exists and both the spin 
and string correlations decay algebraically with gapless excitations.
We also showed that the gapped chiral phase existed in a very narrow region 
of $j$ between the Haldane and the gapless chiral phases.
As $j$ increases, two transitions, the Haldane-gapped chiral transition 
at $j = j_{\rm c1}(\Delta = 0) = 0.473 \pm 0.001$ and 
the gapped chiral-gapless chiral transition at 
$j = j_{\rm c2}(\Delta = 0) = 0.490^{+0.010}_{-0.005}$, occur successively.

In the present calculation, we find that the gapped chiral phase continues 
to exist for $\Delta > 0$.
As an example, we show in Fig. \ref{fig:cor04} the $r$-dependence of 
the spin, chiral, and string correlations for $\Delta = 0.4$ 
and several typical values of $j$ on $\log$-$\log$ plots.
One can clearly see from the figure, both the chiral and string LRO's coexist 
at $j = 0.590$, which indicates that the point $j = 0.590$ belongs to 
the gapped chiral phase.
Performing the calculations with increasing $j$ at intervals of 0.002, 
we estimate the transition points for $\Delta = 0.4$ to be 
$j_{\rm c1} = 0.582 \pm 0.002$ and $j_{\rm c2} = 0.600 \pm 0.003$.

Since $j_{\rm c1}$ lies so close to $j_{\rm c2}$, one may wonder if 
the appearance of the gapped chiral phase might be just an artifact 
of the numerical calculation.
We, however, believe that we can refute such suspicion.
In our infinite-system DMRG, the only source of uncertainty comes 
from the truncation error due to a finite $m$.
As can be seen in Fig. \ref{fig:cor04}, both the chiral correlations in 
the Haldane phase ($j \lsim 0.580$) and the string correlations 
in the gapless chiral phase ($j \gsim 0.600$) tend to be enhanced with 
increasing $m$.
This means that the estimated transition point $j_{\rm c1}$ ($j_{\rm c2}$) 
monotonously decreases (increases) as $m$ is increased, i.e., 
the DMRG calculation tends to underestimate the range of $j$ for 
the gapped chiral phase.
We can therefore safely conclude that the two transition points locate 
very near, but are distinctly different from each other, 
and the gapped chiral phase does exist.
Analyzing the data for several $\Delta$, we find the gapped chiral phase 
in the range $0 \le \Delta \le 0.8$.
The estimated values of $j_{\rm c1}(\Delta)$ and $j_{\rm c2}(\Delta)$ are 
summarized in Table \ref{tab:jc12}.
At $\Delta = 0.9$, no region belonging to the gapped chiral phase 
is observed within the calculation with varying $j$ at intervals of $0.002$.
We, however, cannot precisely determine the point beyond which 
the gapped chiral phase vanishes, 
since we cannot exclude the possibility that the region of the gapped chiral 
phase is just too narrow to detect even at $\Delta = 0.9$.

In order to investigate the critical properties of 
the Haldane-gapped chiral phase transition at $j = j_{\rm c1}$, 
we analyze the chiral correlation length $\xi_\kappa$ in the Haldane phase.
The chiral correlation length is estimated by fitting the chiral correlation 
functions in the range $30 \lsim r \le 100$ to the standard form of 
an exponential decay,
\begin{equation}
C_\kappa (r) = A~ \exp(-r/\xi_\kappa),
\end{equation}
where $A$ is a constant.
The $j$-dependence of the estimated chiral correlation length 
$\xi_\kappa$ in the pure $XY$ case ($\Delta = 0$) 
is shown in Fig. \ref{fig:xi-chl} on a $\log$-$\log$ plot.
It shows a linear behavior near the critical point suggesting 
the standard power-law divergence, 
$\xi_\kappa \sim |j - j_{\rm c1}|^{-\nu_\kappa}$.
By least-square fitting using the three data points near the critical point, 
we estimate $\nu_\kappa = 0.9 \pm 0.1$.
The error is rather large since the exponent is sensitive to the change 
of the critical point $j_{\rm c1}$.
The estimated value of $\nu_\kappa$ is slightly smaller than, 
but consistent with, that of the 2D Ising model, $\nu_\kappa = 1$.

At the gapped chiral-gapless chiral transition at $j = j_{\rm c2}$, 
we extract the exponents of algebraic decay of the spin correlation function 
$C_{\rm s}^x(r)$ in the gapless chiral phase by fitting $C_{\rm s}^x(r)$
in the range $10 \lsim r \le 100$ to the power-law form,
\begin{equation}
C_{\rm s}^x(r) = B~ r^{-\eta_x},
\end{equation}
where $B$ is a constant.
Figure \ref{fig:KT} shows the $j$-dependence of the estimated exponent 
$\eta_x$ in the pure $XY$ case ($\Delta = 0$).
As can be seen in the figure, the exponent $\eta_x$ increases from 
$\eta_x \sim 0.15$ at $j \gsim 0.8$ to $\eta_x \sim 0.29$ at $j = 0.490$ 
as $j$ approaches the critical value, 
$j = j_{\rm c2} = 0.490^{+0.010}_{-0.005}$.
We estimate $\eta_x$ at the critical point to be 
$\eta_x(j=j_{\rm c2}) = 0.28 \pm 0.03 $.
The large error is due to the rather large error in the 
estimated value of $j_{\rm c2}$.
Although the estimated value of $\eta_x$ at $j=j_{\rm c2}$ is not inconsistent 
with the value for the Kosterlitz-Thouless transition, 
$\eta_x = 1/4$, 
more detailed study is needed to identify the universality class of 
the transition between the gapped and gapless chiral phases.

Next, we consider the transition between the chiral and  DH phases 
at $j = j_{\rm c3}$.
As mentioned in {\S} \ref{sec:int}, the system in the DH phase can be 
regarded as two Haldane subchains with the exchange coupling $J_2$ 
coupled with the weak inter-subchain coupling $J_1$.
Thus it is natural to expect the string correlations on each subchain 
to be the order parameter of this phase.
This generalized string correlation, however, does not exhibit a LRO 
in this phase, but decays exponentially\cite{Kolezk1,Kolezk2}.
Consequently, we have no order parameter at hand to characterize the DH phase.
We accordingly determine the phase boundary between the gapless chiral and 
DH phases as the line where the chiral LRO vanishes.
The data of the chiral correlation for $j = 1.0$ and several typical values 
of $\Delta$ are shown in Fig. \ref{fig:chl-DH}.
It is clearly shown in the figure that the system exhibits a chiral LRO 
for $\Delta \lsim 0.68$ while $C_\kappa(r)$ decays exponentially for 
$\Delta \ge 0.70$.
Hence, we estimate the transition point, 
$j_{\rm c3}(\Delta =0.69 \pm 0.01) = 1.0$.
As can be seen in Fig.\ref{fig:diag}, the value of $j_{\rm c3}(\Delta)$ 
estimated in this way becomes larger as $\Delta$ becomes smaller.
Here we note that there is a possibility that the gapped chiral phase also 
exists in the vicinity of the line, i.e., the line where the chiral LRO 
vanishes might be different from the line where the energy gap vanishes.
It is, however, very difficult to distinguish these lines each other, if any, 
since we have no means to estimate the energy gap 
accurately enough to do this.
For that purpose, it may be necessary to introduce another order parameter 
which characterizes the DH phase.

Since the $z$-component of the exchange couplings in the Hamiltonian, 
$S_l^z S_{l+\rho}^z$, reduces the energy barrier between the two 
discretely degenerate states according to the right- and left-handed 
chirality, we can expect naively that the tendency towards chiral ordering 
is suppressed by the introduction of the $z$-component.
This feature can clearly be seen in Fig.\ref{fig:diag}, i.e., 
the chiral phases appear in the narrower range of $j$ 
as $\Delta$ becomes larger.
While the chiral LRO is found for $0 \le \Delta \le 0.9$ in some range of $j$,
we observe no chiral LRO for $\Delta = 0.95$ and $1.0$.
From this observation, we conclude that, between $0.9 < \Delta < 0.95$,
there is a multi-critical point among the Haldane, gapless chiral, and 
DH phases.

\subsection{Haldane-DH transition}
Here we consider the transition between the Haldane and DH phases 
at $j = j_{\rm c4}$.
This transition has been studied by Kolezhuk {\it et al.} 
in the Heisenberg case ($\Delta = 1$)\cite{Kolezk1,Kolezk2}.
From the numerical calculation of the string correlation function 
$C_{\rm str}(r)$ by the DMRG method, they estimated the 
transition point, $j = j_{\rm c4} = 0.7444(6)$, where the string LRO vanished.
They also concluded that the transition was first-order from the observation 
of a discontinuous jump of the string order parameter from a finite value 
to zero.
Our results are consistent with theirs; we have also observed the jump in 
the string order parameter at the transition, although the estimated 
transition point $ j = j_{\rm c4} = 0.747 \pm 0.001$ is slightly different 
from theirs.
The reason of this small difference may be that the number of kept states, 
$m$, of our calculation are larger than that of their calculation.
The first-order nature of the phase transition persists even away from the 
Heisenberg line ($\Delta = 1$).
In Fig. \ref{fig:1st}, we show the $j$-dependence of the string 
correlation at $r \to \infty$ for $\Delta = 0.95$ and $1.20$.
Clearly, the string order parameter disappears discontinuously 
at the transition point.
The estimated transition line runs from the multi-critical point to 
the N\'{e}el-DN phase transition point, $j = 0.5$ in the 
Ising limit $\Delta \to \infty$, which will be explained 
in the next subsection.

\subsection{Transition to the magnetic ordered phases}

In this subsection, we discuss the results for the Haldane-N\'{e}el 
transition at $j = j_{\rm c5}$ and the region between the DH and DN phases.
Some information is available at some special points in the phase diagram.
In the Ising limit, $\Delta \to \infty$, a phase transition 
between the N\'{e}el and the DN phases occurs at $j = 0.5$.
The system exhibits a magnetic LRO in both phases while the nature of 
two-spin correlations changes from the N\'{e}el type ($j \le 0.5$) to 
the ^^ uudd' type ($j > 0.5$).
In the case of $j = 0$, it has also been shown analytically
\cite{Hal-Nel1a,Hal-Nel1b} 
and numerically\cite{Hal-Nel2a,Hal-Nel2b} that there is a phase transition 
between the Haldane and N\'{e}el phases 
at $\Delta = \Delta_{\rm c} \sim 1.18$.
This transition is continuous and belongs to the universality 
class of the 2D Ising model.
In the limit of $j \to \infty$, since the system can be regarded as two 
decoupled chains with the exchange coupling $J_2$, there must be a phase 
transition of the same universality class between the DH and DN phases 
at $\Delta = \Delta_{\rm c}$.

We estimate numerically the phase boundary between the Haldane and 
N\'{e}el phases by examining the existence of the magnetic LRO 
of the N\'{e}el type.
As an example, we show in Fig. \ref{fig:H-N} our numerical results 
of $C_{\rm s}^z (r)$ for $\Delta = 1.5$ and several typical values of $j$, 
which suggests that the transition occurs at 
$j = j_{\rm c5}(\Delta = 1.5) \sim 0.1845$.
As can be seen in Fig. \ref{fig:diag}, the phase boundary determined 
in this way smoothly connects the transition point in the limiting case, 
$\Delta = \Delta_{\rm c}$ for $j = 0$, to the N\'{e}el-DN transition point, 
$j = 0.5$ for $\Delta \to \infty$.
Since the Haldane-N\'{e}el transition at $j = 0$ belongs to 
the universality class of the 2D Ising type, we expect that 
the transition in the intermediate region also exhibits 
the same critical properties.
We analyze the critical behavior of the phase transition 
in the intermediate region at $\Delta = 1.5$.
From the analysis of the order parameter and the correlation length 
associated with the $z$-component of the spin shown in 
Figs.\ref{fig:criH-N}(a) and \ref{fig:criH-N}(b), 
we obtain the critical exponents, $\beta = 0.122 \pm 0.010$ 
and $\nu = 0.96 \pm 0.08$.
We also estimate the critical-point decay exponent $\eta = 0.26 \pm 0.02$ 
from the data of $C_{\rm s}^z(r)$ shown in Fig. \ref{fig:H-N}.
These values of critical exponents are in good agreement 
with the values of the 2D Ising model, $\beta = 1/8, \eta = 1/4$ 
and $\nu = 1$.

Next we consider the region between the DH and DN phases.
As an example, we show in Fig. \ref{fig:DH-DN} the numerical data 
of $C_{\rm s}^z(r)$ for $j = 1.0$ and several typical values of $\Delta$ 
on a semi-$\log$ plot.
In the DH phase ($\Delta \le 1.6$), the spin correlation turns out to be 
incommensurate (note that $C_{\rm s}^z(r)$ in this phase has been divided by 
${\rm cos}(Qr)$ in the figure) and decays exponentially, while it exhibits 
a magnetic LRO of the ^^ uudd' type in the DN phase ($\Delta \ge 2.0$).
In the DH phase, the correlation length associated with the incommensurate 
short-range order becomes longer as $\Delta$ increases.
For $\Delta = 1.7$, $1.8$, and $1.9$, on the other hand, 
our infinite-system DMRG calculation does not converge, i.e.,
the bond energy at the center of the chain fluctuates irregularly 
even after thousands of steps of the algorithm.
This mean that in the region it is impossible to obtain the bulk 
properties of the system within the framework of 
the infinite-system DMRG method.
The origin of this difficulty can be attributed to an approximation 
employed in the algorithm.
As noted in {\S} \ref{sec:DMRG}, the size of the system treated in the 
infinite-system DMRG increases by two at every step.
Accordingly, the set of $m$-states kept in a step is, strictly speaking, 
not the optimal one for the block states in the next step.
In the absence of magnetic LRO or in the presence of 
commensurate LRO with the periodicity of two or four lattice sites 
($q = \pi$ or $\pi/2$), this error of the algorithm becomes 
less and less significant as the system becomes larger, while 
in the presence of incommensurate (or highercommensurate) magnetic LRO 
with a wavenumber $q = Q \ne \pi$, $\pi/2$, 
the error remains even in the thermodynamic limit 
and causes a crucial problem in the calculation\cite{spring}.
We therefore consider that the breakdown of our calculation in the range 
$1.65 \lsim \Delta \lsim 1.95$ suggests that the system has an 
incommensurate magnetic LRO in this region.
From this conjecture, we believe that the point where the magnetic LRO 
of the ^^ uudd' type starts to appear, $j=1.0$ and $\Delta \sim 1.95$ 
(see Fig. \ref{fig:DH-DN}), is a boundary point of the DN phase.
Connecting the points estimated in this way, we determine the 
boundary line of the DN phase (the dashed line in Fig. \ref{fig:diag}).
In contrast, we cannot precisely determine the boundary line of the DH phase 
nor directly verify the existence of the phase with an incommensurate 
magnetic LRO.

\section{Summary}\label{sec:dis}

In this paper we have studied the ground-state properties of 
spin-$1$ $XXZ$ chains with the antiferromagnetic nearest-neighbor and 
the frustrating next-nearest-neighbor couplings.
We have performed the infinite-system DMRG calculation to numerically compute 
the spin, chiral, and string correlation functions.
Analyzing the $r$-dependence of these correlations, we have determined 
the ground-state phase diagram which turns out to include the six phases, 
the Haldane, gapped and gapless chiral, DH, N\'{e}el, and DN phases 
(Fig.\ref{fig:diag}).
The chiral phases appear in a quite broad region in the phase diagram for 
$\Delta < 0.95$, while the region becomes narrower as $\Delta$ increases.
We have also confirmed that the gapped chiral phase does exist in a very 
narrow but finite region between the Haldane and gapless chiral phases.
The multi-critical point among the Haldane, gapless 
chiral, and DH phases exists between $0.9 < \Delta < 0.95$.
Furthermore, we have investigated the critical properties of 
the phase transitions.
For the Haldane-gapped chiral transition, the chiral correlation-length 
exponent is estimated to be $\nu_\kappa = 0.9 \pm 0.1$, 
which is not inconsistent with the 2D Ising value, $\nu_\kappa = 1$.

The quantum chiral phases which we have discovered in this and previous work 
are novel ones in quantum spin systems.
Accordingly, many interesting open problems about these phases still remain.
As mentioned in {\S} \ref{sec:int}, in the $S = 1$ case only the chirality 
exhibits a LRO without the helical LRO in the chiral phases 
while both LRO coexist in the helical ordered phase 
which appears in the classical case, $S \to \infty$.
This means that the quantum fluctuation in the $S = 1$ chain recovers 
only the time-reversal symmetry with the parity symmetry kept broken.
In the $S = 1/2$ case, strong quantum fluctuation eventually recovers both 
symmetries.
In order to gain more insight into the effect of quantum fluctuation 
on each symmetry, it is necessary to investigate how the phase diagram 
and the nature of the chiral phases depend on the spin quantum number $S$, 
particularly paying attention to the possible difference between 
integer and half-odd integer $S$.
We are now performing calculations in the $S = 3/2$ case.
A preliminary calculation suggests that the gapless chiral phase 
appears in this case as well.

\section*{Acknowledgments}
We thank K. Nomura and T. Momoi for useful discussions.
Numerical calculations were carried out in part at the Yukawa Institute 
Computer Facility, Kyoto University.
T. H. was supported by a Grant-in-Aid for Encouragement of Young Scientists 
from Ministry of Education, Science and Culture of Japan.

\vspace{3.0cm}
\begin{table}
\caption{The long-distance behavior of the spin, chiral, and string 
correlations in each phase.
"expo." and "power" means exponential decay and power-law decay, 
respectively.} 
\label{tab:cor}
\begin{tabular}{lcccccc}
             &Haldane  &gapped chiral&gapless chiral& DH  & N\'{e}el& DN   \\
\hline
$C_{\rm s}^z$&  expo.  &   expo.     &     power    &expo.&  LRO    & LRO  \\
$C_{\rm s}^x$&  expo.  &   expo.     &     power    &expo.&  expo.  & expo.\\
$C_\kappa$   &  expo.  &   LRO       &     LRO      &expo.&  expo.  & expo.\\
$C_{\rm str}$&  LRO    &   LRO       &     power    &expo.&  LRO    & LRO  \\
\end{tabular}
\end{table}%

\begin{table}
\caption{The estimated values of $j_{\rm c1}(\Delta)$ and $j_{\rm c2}(\Delta)$
for various values of $\Delta$.} 
\label{tab:jc12}
\begin{tabular}{cccccc}
$\Delta$  &  0                       &   0.2             &   0.4             &  0.6    &  0.8    \\
\hline
$j_{\rm c1}$  & $0.473 \pm 0.001$        & $0.527 \pm 0.002$ & 
$0.582 \pm 0.002$ & $0.642 \pm 0.002$ & $0.715 \pm 0.003$  \\
$j_{\rm c2}$  & $0.490^{+0.010}_{-0.005}$ & $0.540 \pm 0.003$ &
$0.600 \pm 0.003$ & $0.660 \pm 0.003$ & $0.720 \pm 0.002$
\end{tabular}
\end{table}%

\newpage
\begin{figure}
   \caption{The ground-state phase diagram. The circles, squares, 
   and diamonds represent the points where the string LRO, chiral LRO, 
   and magnetic LRO vanishes, respectively.}
  \label{fig:diag}
\end{figure}%

\begin{figure}
   \caption{The $r$-dependence of the correlation functions 
   for $\Delta = 0.4$: 
   (a) spin correlation $C_{\rm s}^x(r)$ divided 
   by the oscillating factor ${\rm cos}(Qr)$; 
   (b) chiral correlation $C_\kappa(r)$; 
   (c) string correlation $-C_{\rm str}(r)$.
   The number of kept states is $m = 260$ (for $j = 0.570$), $m = 300$ 
   (for $j = 0.580, 0.590, 0.610$) and $m = 400$ (for $j = 0.600$).
   To illustrate the $m$-dependence, we also indicate the data 
   with smaller $m$ by crosses for several cases: 
   in figure (b), $m = 180, 220$ for $j = 0.570$ and $m = 220, 260$ 
   for $j = 0.580$;
   in figure (c), $m = 300, 350$ for $j = 0.600$ and $m = 260, 300$ 
   for $j = 0.610$.
   In the other cases, the numerical errors of the data are smaller 
   than the symbols.
   Arrows in the figures represent the extrapolated $r=\infty$ values.}
  \label{fig:cor04}
\end{figure}%

\begin{figure}
   \caption{The $\log$-$\log$ plot of the $j$-dependence of the chiral 
   correlation length for $\Delta = 0$ ($XY$ case) 
   in the critical region of the Haldane phase ($j < j_{\rm c1}$).
   We plot the data setting $j_{\rm c1} = 0.4729$.}
  \label{fig:xi-chl}
\end{figure}%

\begin{figure}
   \caption{The $j$-dependence of the exponent of algebraic decay of 
   $C_{\rm s}^x(r)$ for $\Delta = 0$
   in the gapless chiral phase ($j > j_{\rm c2} \sim 0.490$).
   The vertical dashed line represents the phase boundary $j_{\rm c2}$.}
  \label{fig:KT}
\end{figure}%

\begin{figure}
   \caption{The $r$-dependence of the chiral correlation function 
   for $j = 1.0$ on a $\log$-$\log$ plot.
   The number of kept states is $m = 300$ (for $\Delta = 0.60, 0.65, 0.72$), 
   $m = 350$ (for $\Delta = 0.68$) and $m = 400$ (for $\Delta = 0.70$).
   For $\Delta = 0.68$, where the $m$-dependence of the data is rather large, 
   we indicate the data with smaller $m$ by crosses ($m = 260, 300$).
   For the other $\Delta$, the numerical errors of the data are smaller 
   than the symbols.
   Arrows in the figures represent the extrapolated $r=\infty$ values.}
  \label{fig:chl-DH}
\end{figure}

\begin{figure}
   \caption{The strength of the string correlation function at $r \to \infty$
   as a function of $j$ for $\Delta = 0.95$ (solid circles) and 
   $1.20$ (open circles). The vertical dashed lines represent 
   the transition points $j_{\rm c4}$ for each $\Delta$.}
  \label{fig:1st}
\end{figure}%

\begin{figure}
   \caption{The $r$-dependence of the spin correlation function 
   $C_{\rm s}^z(r)$ for $\Delta = 1.5$ on a $\log$-$\log$ plot.
   The number of kept states is $m = 220$ for all $j$.
   The numerical errors of the data are smaller than the symbols.
   The solid line represents the power-law decay of the correlation 
   at $j = 0.1845$.}
  \label{fig:H-N}
\end{figure}%

\begin{figure}
   \caption{(a) The strength of the spin correlation function 
   $C_{\rm s}^z(r)$ at $r \to \infty$ in the N\'{e}el phase 
   ($j < j_{\rm c5}$) as a function of $j$ 
   for $\Delta = 1.5$.
   We plot the data setting $j_{\rm c5} = 0.1845$.
   We extract the exponent $\beta$ by fitting the data to the form, 
   $C_{\rm s}^z(r \to \infty) \sim |j - j_{\rm c5}|^{2\beta}$.
   (b) The correlation length associated with $C_{\rm s}^z$ 
   in the Haldane phase ($j > j_{\rm c5}$)
   as a function of $j$ for $\Delta = 1.5$.
   We plot the data setting $j_{\rm c5} = 0.183$.
   Fitting the data to the form, $\xi \sim |j - j_{\rm c5}|^{\nu}$, 
   we estimate the exponent $\nu$.}
  \label{fig:criH-N}
\end{figure}%

\begin{figure}
   \caption{The $r$-dependence of the spin correlation function 
   $C_{\rm s}^z(r)$ for $j = 1.0$ on a semi-log plot.
   For $\Delta \le 1.6$ (the DH phase), $C_{\rm s}^z(r)$ is divided 
   by the oscillating factor $f(r) = {\rm cos}(Qr)$, while it is divided 
   by $f(r) = \exp({\rm i}\pi [r/2])$ for $\Delta \ge 2.0$ (the DN phase).
   The number of kept states is $m = 180$ (for $\Delta = 2.0, 2.5$), 
   and $m = 260$ (for $\Delta = 1.2, 1.4, 1.6$).
   The numerical errors of the data are smaller than the symbols.
   Arrows in the figures represent the extrapolated $r=\infty$ values.}
  \label{fig:DH-DN}
\end{figure}%

\end{document}